# Energy, cost, and $CO_2$ emission comparison between radiant wall panel systems and radiator systems

**Milorad Bojić[1], Dragan Cvetković[1,*], Marko Miletić[1], Jovan Malešević[1], Harry Boyer[2]**

[1]*Faculty of Engineering Science, University of Kragujevac, 34000 Kragujevac, Serbia*

[2]*Department PIMENT Lab., University of Réunion Island, France*

**Abstract:**

The main goal of this paper is to evaluate the possibility of application or replacement of radiators with low-temperature radiant panels. This paper shows the comparison results of operations of 4 space heating systems: the low-temperature radiant panel system without any additional thermal insulation of external walls (PH-WOI), the low-temperature radiant panel system with additional thermal insulation of external walls (PH-WI), the radiator system without any additional thermal insulation of external walls (the classical heating system) (RH-WOI), and the radiator system with additional thermal insulation of external walls (RH-WI). The operation of each system is simulated by software EnergyPlus. The investigation shows that the PH-WI gives the best results. The RH-WOI has the largest energy consumption, and the largest pollutant emission. However, the PH-WI requires the highest investment.

**Nomenclature**

$A$ = area of heat emitters, $m^2$

$E$ = energy, J/year

$f$ = specific cost, €/$m^2$

$g$ = $CO_2$ emission factor, kg/J

$k$ = correction coefficient of the natural gas consumption

$L$ = lifecycle, year

$l$ = specific energy, J/kg

$\dot{m}$ = mass flow rate of water, kg/s

$m_1$ = fixed monthly cost for metering, Euro

$q$ = heat flux, W

$R$ = primary energy consumption coefficient,

$S$ = carbon dioxide emission, kg/year

$T$ = temperature, K

$U$ = coefficient of heat transfer, W/($m^2$K)

$V$ = operating cost, Euro

$\delta$ = thickness, m

$\varepsilon$ = efficiency of the heat transfer from the radiant wall panel to the air in the heated room,

$\rho$ = density, kg/$m^3$

**Subscripts and superscripts**

a = air,

*a* = average,

*Emb* = embodied,

*el* = electricity,

*i* = number of zone,

*in* = inlet,

*ng* = natural gas,

*o* = outside,

*sys* = system,

*TOT* = total,

w = water,

**Abbreviations**

PH-WOI = panel heating system without additional thermal insulation

PH-WI = panel heating system with additional thermal insulation

RH-WOI = radiator heating system without additional thermal insulation

RH-WI = radiator heating system with additional thermal insulation

SHGC = solar heat gain coefficient

UFAD = underfloor air distribution

## 1. Introduction

In Serbia, the concept of the radiant panel heating system (wall and floor panel) is relatively well-known, but its application is still in its infancy. This is probably as a result of insufficient information. However, the introduction of new materials, better control strategies, and better utilization of alternative energy sources increase their demand and importance in the HVAC industry. To prove the usefulness of such systems, it is necessary to compare these heating systems with the classical heating systems regarding their energy, economy, and environmental influence.

In the literature, there are many papers dealing with investigations of the low temperature radiant systems and their comparison with other heating systems regarding energy consumption and obtained thermal comfort. Chen [1] compares the thermal comfort and the energy consumption among a ceiling radiant heating system, a radiator heating system and a warm air heating system by using airflow program Phoenics-84 and air-conditioning load program Accuracy. Haddad et al. [2] compares the performance of a forced-air and a radiant floor residential heating system connected to solar collectors. Raftery et al. [3] compares the performance of a joint underfloor air distribution (UFAD) and radiant hydronic system with a typical only UFAD system. Imanari et al. [4] compares a ceiling panel heating system in conjunction with an air-conditioning system and a conventional air-conditioning system in term of thermal comfort, energy efficiency and cost efficiency. Stetiu [5] shows that using a radiant cooling system instead of a traditional all-air system in office space could save on average 30% of the energy consumption. Nobody compared the radiant wall heating system with other heating systems in terms of energy, economy, and environmental influence.

The basic aim of this paper is to compare the radiant wall heating system with the radiator heating system in terms of energy, economy, and environmental influence. During this research, the operations of four heating systems are simulated such as: PH-WOI, PH-WI, RH-WOI, and RH-WI. For one heating season, the consumption of natural gas for heating, the consumption of electricity for heating, and the energy costs are analyzed as simulation results. In addition the carbon dioxide emission is analyzed. When analyzing the energy consumption and the pollutant emission, the embodied energy and the embodied $CO_2$ of additional thermal insulation are considered. Furthermore, the investment into these systems is compared.

## 2. Building

### 2.1 Building description

An investigated residential building is shown in Fig.1. The building has two stories with 20 rooms; 18 of them are heated with the heating bodies located inside these rooms. The rooms that are not heated are the storage room at the first floor designated as Pr 8 and the room at the second floor designated as Sp 10. Other rooms are heated according the required temperature schedule [6]. The design temperature for indoor air in the living room, bedroom, and kitchen is 20 °C, and in the hallway, and stairs 18 °C.

It is assumed that this building is not surrounded by any other nearby buildings.

*<Fig.1 Isometric appearance of analyzed residential building with cuts for easier understanding of distribution and appearance of rooms  Here, Pr 1=hallway, Pr 2=stair, Pr 3=room, Pr 4= bathroom, Pr 5=bedroom, Pr 6=living room, Pr 7=kitchen, Pr 8=storage, Pr 9=toilet, Pr 10= bathroom, Sp 1=hallway, Sp 2=stair, Sp 3=bedroom, Sp 4=bathroom, Sp 5=toilet, Sp 6=working room, Sp 7=living room, Sp 8=kitchen, Sp 9=bedroom, and Sp 10=passage.>*

The total floor area of the building is 301 $m^2$. The gross external wall area is 306 $m^2$, and the gross roof area is 187 $m^2$. The external wall of the building consists of the thermal insulation (Styrofoam) of 50mm thickness, the clay block (hollow brick) of 190mm thickness and the lime mortar of 20mm thickness. The U-value of this wall is 0.57W/($m^2$K).  The total window glazing area is 27 $m^2$. The percentage of the glazing compared to the area of the overall external wall is 8.76%. The windows are double glazed with U-value=2.72 W/($m^2$K), and SHGC=0.764.

The building loses heat through its envelope by conduction, and by air infiltration at 0.5 air changes per hour.

The investigated building is used by two households with the total of 8 people. As electrical consumers in the building, the classical electrical household devices are used (stoves, refrigerators, TVs, radios, washing machines, freezers, and microwaves). In the building, the traditional hanging lamps are used.

## 2.2 Additional thermal insulation plate from wood wool

To find better technical solution for application of radiant wall panel heating, it is necessary to use some heat transfer barrier such as a thermal insulation plate. This heat barrier stops heat dissipation from the radiant wall panel to the outdoors. However, it directs the heat to the heated room. These plates represent the pressed mixture of long wood fibres and magnesite. As the final product, hard and stable plates are obtained. The interior of these plates contains many air voids. Around 70% of the plate volume presents the air voids that excellently stop the heat transfer through the wall. Because of the air voids, these plates are also resistant to the negative influence of humidity. The characteristic of these plates are given in Table 1.

*<Table 1.The characteristic of the thermal insulation plates from wood wool [7]>*

# 3. Mathematical model

## 3.1 Used Software

To simulate a thermal behaviour of the building and have accurate calculation results, software EnergyPlus is used. This program is a very useful tool for modelling of energy and environmental behaviour of buildings. The program is initially developed by Lawrence Berkeley National Laboratory, U.S. Army Construction Engineering Laboratory, and the University of Illinois [8]. In the software, it is possible to input how people use building during its space heating. In this direction, the complex schedules of heating can be defined together with the schedules for use of lighting, internal energy devices and occupancy in the building. The influence of the solar radiation, shadowing and infiltration is also taken into consideration [6, 9]. The investigated building and its heating systems are modelled by using this software. Then, the general components of these systems are modelled such as the boiler, pumps, convective baseboard heaters, and radiant wall panel heaters.

## 3.2 Heating systems

The investigated heating systems operate in the building located in Belgrade, Serbia. They operate by using hot water at different temperatures. The radiant wall panel heating operates with low water temperatures. The classic non-condensing boilers used to generate heat, for all heating systems by using natural gas. The boiler appears as electric consumers in the system. The parasitic electric load from the boiler is consumed whenever the boiler is operating. The model assumes that this parasitic power doesn't contribute to the heating of the water. Its load is about 75W when it operates. The water circulation pump also uses electricity to operate. This is taken into account to calculate the primary energy consumption.

### 3.2.1 Radiator heating system

A radiator heating system represents an installation of the central heating. In the system, heat is generated in the boiler. For the generation of the heat, the natural gas boiler is used where the chemical energy of natural gas is transferred into the heat. Then, the heat is distributed by hot water (heat carrier) to the radiators. The radiators heat the rooms. The radiators are installed in each heated room of the house. The hot-water is circulated by a water circulation pump, which operates continuously. If the valves stop the hot-water flow, then this flow goes through the bypass pipe. The radiators, as rule of thumb, are located next to the cold surfaces of the envelope. Then, they significantly influence the thermal comfort. The radiators release the highest amount of heat to the heated room by convection and one part by heat radiation [10]. To simplify the problem, the radiator heating is modelled in EnergyPlus software by the convector model. EnergyPlus software would define the size of the radiators based on the heat behaviour of the building. The equation of heat flux between the radiator water and the room air is given as [6]:

$$q=UA(T_{a,w}-T_{a,a}) \quad (1)$$

Here, $T_{a,w}$,$T_{a,a}$ stand for the mean temperatures of the radiator water and the room air, respectively, *UA* stands for the product of the coefficient of heat transfer, and the heat transfer area.

In this system, $T_{a,w}$ is set at 70 °C. The inlet water temperature to the radiators is set at 80 °C.

### 3.2.2 Radiant wall panel heating systems

Radiant wall heating panels occupy large wall surfaces. In the investigated house, the heating panels cover all external walls of the heated rooms - their surface area is the same as that of these walls (212 $m^2$). The heating panels are not used in the house floors and the house ceilings.

The radiant wall heating panels contain the inner pipes with the hot water. They may exchange up to 70% of the heat with the heated room by thermal radiation [10]. Because of low temperature of the heating panel and its hot water, the heat transfer rate from the heating surface is relatively small to the heated room. This heating system consists of the radiant wall heating panels, the water circulation pump, valves, and the natural gas boiler. The boiler operates with relatively low temperature in the supply pipe and with small temperature difference between the supply water and the return water. Through the entire heating system, the hot water is circulated by a water circulation pump. The pump operates continuously and if valves stop the hot-water flow, then the flow goes through the bypass pipe.

Because of high share of the radiation from the radiant wall panel heating system, a feeling of the thermal comfort is obtained at much lower air temperature in the room than that by using other heating systems.

The release of heat from the radiant wall panel depends on many factors such as the type of wall covering, the mutual distance of pipes in the panel, the dimension of pipes, the structure of the wall, and the type of radiant wall panel system.

EnergyPlus software is used for the simulation of the application of the radiant wall panel heating systems as the software has all imbedded functions that are necessary for the successful simulation of this application in buildings. The software uses the basic equation of heat balance for the internal partitions (walls, ceilings, and floors), the envelope (walls, floors, and roofs) and the air in different zones of the building. The transfer of heat through each of constructions is modelled by using the conduction heat transfer functions [9]. The equation that defines the heat flux to/from the heating panel [6]:

$$q = \frac{T_{w,in} - C_3 - C_1 C_5}{\frac{1}{\varepsilon (\dot{m} c_p)_{water}} + C_4 + C_2 C_5} \qquad (2)$$

where $T_{w,in}$ stands for the inlet temperature of the hot water in the radiant wall heating panel, $\dot{m}$ stands for the mass flow rate of the hot water inside the radiant wall heating panel, $c_p$ stands for the specific heat capacity of the hot water, $\varepsilon$ stands for the efficiency of the heat transfer from the radiant wall heating panel to the air in the heated room, $C_1$ stands for the coefficient that includes surface heat balance and past history terms as well as the influence of the current outside temperature, $C_2$ stands for the coefficient that depends on the heat source transfer function term and the coefficients of terms linked directly to the inside surface temperature, $C_3$ stands for the coefficient that includes all of the history terms and the effect of the current outside temperature, $C_4$ stands for the heat source transfer function for the current time step, $C_5$ stands for the conduction transfer function for the inside surface temperature at the current time step. These coefficients and other variables (the exit temperature of the hot water, and the external and internal temperature of the wall surface) may be calculated by using equations given in [6].

In this system, the mean temperature of the hot water in the heating panels is set at 37 °C. The inlet water temperature $T_{w,in}$ in the heating panels is set at 40 °C.

### 3.3 Weather

The investigated building is located in Belgrade. Belgrade is a capital city of Serbia. Its average height above the sea level is 132m. Its latitude is 44°48N, and longitude 20°28E. The city has a moderate continental climate with distinct seasons (winter, spring, summer, and autumn). The weather file used in the simulation is obtained by measurements at the Belgrade weather station [11].

## 3.4 Total energy consumption of system

As the results, the simulations by EnergyPlus give the energy consumption from natural gas per heating season ($E_{ng}$), and the energy consumption from electricity per heating season ($E_{el}$).

### 3.4.1 Primary energy consumption of heating system

The primary energy consumption per heating season at the investigated building is calculated by using the following equation:

$$E_{sys} = E_{ng} + E_{el} R \qquad (3)$$

Here, R stands for the primary energy consumption coefficient. This coefficient is defined as the ratio of the total input energy of energy resources (hydro, coal, oil and natural gas) and the finally generated electric energy. For the Serbian energy mix for electricity production, R = 3.61 [12].

### 3.4.2 Embodied energy

The embodied energy is defined as the commercial energy (fossil fuels, nuclear, etc) that was used in the work to make any product. The embodied energy consumption per heating season for the investigated building is calculated by using the following equation:

$$E_{Emb} = \frac{l_{Emb} \cdot \rho \cdot \delta \cdot A}{L}. \qquad (4)$$

Here, $l_{Emb}$ stands for the specific embodied energy of the additional thermal insulation plate, $\rho$ stands for the specific density of the additional thermal insulation plate, $\delta$ stands for the thickness of the additional thermal insulation plate, $A$ stands for the application area of the additional thermal insulation plate, and $L$ stands for the lifecycle of the heating system. The lifecycle of each heating system is set at 20 years [10]. For the additional thermal insulation plate, its geometrical characteristics, and the values of embodied energy are given in Table 1.

### 3.4.3 The total energy consumption of the system

The total energy consumption of the system is the sum of the primary energy consumption by the heating system and the embodied energy of the additional thermal insulation panel. It is given by the following equation:

$$E_{TOT} = E_{sys} + E_{Emb}. \quad (5)$$

## 3.5 Total operating cost

The total operating cost to run the heating system is calculated by using the following equation

$$V_{TOT} = f_{el}E_{el} + f_{ng}E_{ng}k\, m_1. \quad (9)$$

Here, $f_{\mathrm{el}}$ stands for the specific cost of the consumption of electricity, $f_{\mathrm{ng}}$ stands for the specific cost of consumption of natural gas with the energy value of 33338 kJ/m$^3$, $k$ stands for the correction coefficient of the natural gas consumption $k$=1.068 [13], $m_1$ stands for the fixed monthly cost for gas metering [13]. In this equation, the fixed monthly cost for electricity metering is not included because this cost is not significant. These cost factors are given in Table 2.

*<Table 2.The price of energy in Serbia in May 2011 [14,15]>*

## 3.6 Carbon dioxide emission

### 3.6.1 Carbon dioxide emission of heating system

The carbon dioxide emission of the heating system during its operation is calculated by using the following equation:

$$S_{sys} = g_{ng}E_{ng} + g_{el}E_{el}. \quad (10)$$

Here, $g_{\mathrm{ng}}$ stands for the specific $CO_2$ emission factor for natural gas (kg $CO_2$/GJ), $g_{\mathrm{el}}$ stands for the specific $CO_2$ emission factor for electricity (kg $CO_2$/GJ). The emission factors are given in Table 3.

*<Table 3.The $CO_2$ emission factors for the electric energy and natural gas [7]>*

### 3.6.2 Embodied $CO_2$ emission of additional thermal insulation plate

The embodied $CO_2$ emission of the additional thermal insulation plate is calculated by using the following equation:

$$S_{Emb} = \frac{g_{Emb} \cdot \rho \cdot \delta \cdot A}{L}. \tag{11}$$

Here, $g_{\mathrm{Emb}}$ stands for the specific embodied $CO_2$ emission factor of the additional thermal insulation plate. The emission factors are given in Table 1.

### 3.6.3 Total $CO_2$ emission of the system

The total $CO_2$ emission of the investigated heating system is the sum of the $CO_2$ emission of the heating system and the embodied $CO_2$ emission of the additional thermal insulation plate. It is represented by the following equation:

$$S_{TOT} = S_{sys} + S_{Emb}. \tag{12}$$

## 4. Results and discussion

In these investigations, the four heating systems: the RH-WOI (the traditional system often used in households in Serbia), the RH-WI, the PH-WOI, and the PH-WI are simulated during their operation at the heating season. In entire Serbia, the heating season starts at 15 October and ends at 15 April next year.

### 4.1.1 Results

The simulation gives the results for the energy consumption, operating costs, and $CO_2$ emission for each of all four heating systems installed in the investigated building.

In Fig.2, the consumptions of natural gas energy, electrical energy, primary energy, and total energy are given for each heating system per heating season.

*<Fig. 2 Energy consumption per heating season.>*

As the electricity consumption is very small to be seen in Fig 2, it is shown in Fig 3.

*<Fig. 3 Electric consumption of water circulation pump per heating season.>*

Figure 4 shows the effect of additional thermal insulation to the energy consumption for different heating systems.

*<Fig. 4 Effect of additional thermal insulation to the energy consumption for different heating systems: a) Radiator heating system, b) Panel heating system>*

Figure 5 shows the effect of heating system to the energy consumption.

*<Fig. 5 Effect of heating system to the energy consumption for different walls: a) without additional thermal insulation, b) with additional thermal insulation>*

Figure 6. shows the nominal power of the boilers serving the investigated building for each heating system.

*Fig. 6 Nominal power of boiler.>*

The investment in € in each heating system is shown in Table 4.

*<Table 4. Investments in investigated systems, prices in Euros (€) [16,17]>*

In Fig.7, the operating costs are shown for each simulated heating system. The figure also gives the total operating cost during the heat season, which includes the costs for electricity and natural gas for heating.

*<Fig. 7. Operating cost of heating per heating season.>*

The yearly $CO_2$ emissions are shown for each heating system in Fig. 8. The figure takes into account that the emission factors for Serbia are $g_{el}$ =206.53 kg/GJ, and $g_{ng}$ =56.1 kg/GJ.

*<Fig. 8 CO2 emission per heating season.>*

### 4.1.2 Discussion

By using all presented results in Figs. 2 to 8, the energy, cost, and $CO_2$ emission are analyzed, discussed, and compared for the radiant wall heating panel systems and radiator systems. From Fig.2, it may be found that the consumption of the natural gas for space heating is the lowest for the PH-WI ($E_{ng}$=40.54 GJ/year) and the highest for RH-WOI ($E_{ng}$=66.18 GJ/year) that makes their difference of 39%. From Fig.3, the electrical consumption is the lowest for the PH-WI ($E_{el}$=0.14 GJ/year) and the highest for RH-WOI ($E_{el}$=0.23 GJ/year) that makes their difference of 40%. From Fig.2, the primary energy consumption for space heating is the lowest for the PH-WI ($E_{sys}$=40.96 GJ/year) and the highest for RH-WOI ($E_{sys}$=66.88 GJ/year). Then, the relative difference in $E_{sys}$ between PH-WI and RH-WOI is 39%. Also, the total energy consumption for space heating is the lowest for the PH-WI ($E_{TOT}$=45.03 GJ/year) and the highest for RH-WOI ($E_{TOT}$=66.88 GJ/year). Then, the relative difference in $E_{TOT}$ between PH-WI and RH-WOI is 33%. It can be seen the relative difference in $E_{TOT}$ between PH-WI and RH-WOI is lower than the relative difference in $E_{sys}$ between PH-WI and RH-WOI by 6%. Note that $E_{TOT} = E_{sys} + E_{Emb}$.

Regarding sensitivity to additional thermal insulation, Fig. 4 shows that the panel heating systems are more sensitive to application of additional thermal insulation than the radiator heating systems. With application of the additional thermal insulation in the house with the heating panel systems, the primary energy consumption by the heating panel systems is reduced by 22% and the total energy consumption (when the embodied energy is taken into account) by 14%. Then, the primary energy consumption by the radiator systems is reduced by 15% and the total energy consumption by 8%.

Regarding sensitivity to the used heating system, Fig.5 shows that the radiator heating system consumes more primary energy than the panel heating systems. For the house without additional thermal insulation, the radiator heating system consumes 21% more primary energy than the panel heating systems. Furthermore, for the house with additional thermal insulation, the radiator heating system consumes 28% more primary energy than the panel heating systems, and 26% more total energy (when the embodied energy into thermal insulation is taken into account) than the panel heating system.

From Fig.6, the nominal power of the boiler is the lowest for the PH-WI ($Q_n$=24.9 kW), and the highest for RH-WOI ($Q_n$=27.6 kW). The relative difference in the nominal power of the boiler between RH-WOI and PH-WI is 10%.

From Fig.7, the cost of the primary energy for space heating is the lowest for the PH-WI ($V_{TOT}$=1.76 €/m$^2$), and the highest for RH-WOI ($V_{TOT}$=2.84 €/m$^2$), that makes their difference of 38%. Also, Table 4 shows that the most expensive system is the PH-WI (15079 €), and the most inexpensive system is the RH-WOI (4369 €). The investment in the PH-WI is about 3 times higher than that in the RH-WOI.

From Fig.8, the CO2 emission of the heating systems due to use of energy for the space heating is the lowest for the PH-WI ($S_{sys}$ = 2.30 tCO2/year) and the highest for RH-WOI ($S_{sys}$ =3.76 tCO2/year), that makes their difference of 39%. The total CO2 emission is the lowest for the PH-WI ($S_{TOT}$ =2.50 tCO2/year), and the highest for RH-WOI ($S_{TOT}$ =3.76 tCO2/year) that makes their difference of 33%.

Finally, this investigation of the four heating systems points out that in Serbia the PH-WI is the most energy efficient, and has the lowest negative impact on the environment. However, the PH-WI is the most expensive. This analysis is done with natural gas as an energy source and refers to only the space heating.  In the future, an analysis may be done for space heating and cooling by using low-temperature sources with heat pump [18].

## Conclusion

This paper reports the results of investigations of performances of the four heating systems: RH-WOI, PH-WOI, PH-WI, and RH-WOI. These heating systems operate in the residential house in Belgrade, Serbia. Additionally, this study takes into account the embodied energy and embodied $CO_2$ of the additional thermal insulation of the external building wall. The operation of heating systems is evaluated by analysing their energy consumption, cost, and $CO_2$ emission.

Regarding sensitivity to additional thermal insulation, the panel heating systems are more sensitive to application of additional thermal insulation than the radiator heating systems. Regarding sensitivity to the used heating system, the radiator heating system consumes more primary energy than the panel heating systems. It is found that the PH-WI has the smallest total consumption of energy, and the smallest impact to environment. However, the economic analysis shows that the PH-WI has the smallest cost of space heating but is the most expensive in terms of investment. The additional thermal insulation prevents a significant thermal conduction loss to the outdoors. This impact of the additional thermal insulation is much higher with the panel heating systems than with radiator systems. The embodied energy and embodied $CO_2$ of additional thermal insulation does not have significant adverse impact on heating systems.

## Acknowledgement

This investigation is part of the project TR 33015 of Technological Development of the Republic of Serbia and the project III 42006 of Integral and Interdisciplinary investigations of the Republic of Serbia. We would like to thank to the Ministry of Education and Science of Republic of Serbia for financial support during this investigation.